\documentclass{mn2e}

\usepackage{epsfig}
\usepackage{graphics}
\newcommand{\gsim}{\mathrel{\raise0.35ex\hbox{$\scriptstyle >$}\kern-0.6em 
\lower0.40ex\hbox{{$\scriptstyle \sim$}}}}
\newcommand{\lsim}{\mathrel{\raise0.35ex\hbox{$\scriptstyle <$}\kern-0.6em 
\lower0.40ex\hbox{{$\scriptstyle \sim$}}}}

\def\ang{\mbox{\AA}}
\def\kms{\mbox{${\rm kms}^{-1}$}}
\def\ergsec{\mbox{${\rm erg}\thinspace{\rm s}^{-1}$}}
\def\ergsqcmsec{\mbox{${\rm erg \thinspace cm}^{-2}{\rm s}^{-1}$}}
\def\deg{\hbox{$^\circ$}}

\def\asec{\hbox{$^{"}$}}
\def\msun{\mbox{${\rm M}_{\odot}$}}

\def\lsun{\mbox{${\rm L}_{\odot}$}}
\def\msunyr {\mbox{$M_{\odot}\mbox{yr}^{-1}$}}
\def\ale{\mathrel{\hbox{\rlap{\hbox{\lower4pt\hbox{$\sim$}}}\hbox{$<$}}}}
\def\age{\mathrel{\hbox{\rlap{\hbox{\lower4pt\hbox{$\sim$}}}\hbox{$>$}}}}

\renewcommand{\thefootnote}{\fnsymbol{footnote}}

\begin{document}
\date{\today}

\title[Is NGC 3108 an astronomical hermaphrodite?]{Is NGC 3108 transforming itself from an early to late type galaxy --
an astronomical hermaphrodite?}
\author[Hau et al]{George~K.~T.~Hau$^{1,2}$,  Richard~G.~Bower$^1$, Virginia~Kilborn$^2$, \newauthor
Duncan A. Forbes$^2$, Michael L.~Balogh$^3$, Tom Oosterloo$^4$
 \\
$^1$Physics Department, University of Durham, South Road, Durham DH1 3LE, U.K.\\
$^2$Centre for Astrophysics and Supercomputing, Swinburne University of Technology,
Hawthorn, Victoria, Australia, 3122. \\
$^3$Department of Physics and Astronomy, University of Waterloo, Waterloo, ON, N2L 3G1, Canada. \\
$^4$ASTRON, Oude Hoogeveensedijk 4, 7991 PD Dwingeloo, The Netherlands.
}

\maketitle

\begin{abstract}
A common feature of hierarchical galaxy formation models is the process of
``inverse'' morphological transformation: a bulge dominated galaxy accretes 
a gas disk, dramatically reducing the system's bulge-to-disk mass ratio.
During their formation, present-day galaxies may execute many such
cycles across the Hubble diagram.

A good candidate for such a ``hermaphrodite'' galaxy
 is NGC 3108: a dust-lane early-type galaxy which 
has a large amount of HI gas distributed
in a large scale disk. We present narrow band H$\alpha$ and R-band imaging, 
and compare the results with the HI distribution from the literature. The emission is in two components: a nuclear bar and an extended disk component which coincides with the HI distribution. This suggests that  a stellar disk is currently being formed out of the HI gas. The spatial distributions of the H$\alpha$ and  HI emission and the HII regions are consistent with a barred spiral structure, extending some $20$ kpc in radius.  We measure an extinction-corrected star formation rate of 0.42 \msunyr. The luminosity function of the HII regions is similar to other spiral galaxies, with a power law index of -2.1, suggesting that the star formation mechanism is similar to other spiral galaxies. 

We measured the current disk mass and find that it is too massive to
have been formed by the current star formation rate over the last few
Gyr. It is likely that the star formation rate (SFR) in NGC 3108 was higher in the past. With the current SFR, the  disk in  NGC 3108 will grow to be $\sim 6.2\times 10^9$ \msun\ in stellar mass within the next 5.5 Gyr. While this is substantial, the disk will be insignificant compared with the large bulge mass: the final stellar mass disk-to-bulge ratio will be $\sim 0.02$. NGC 3108 will fail to transform into anything resembling a spiral without a boost in the SFR and additional supply of gas.

\end{abstract}


\section{Introduction}

It is widely accepted that galaxies may change their morphology as they
evolve. Usually, such morphological transformation drives galaxies from
late to early types, often as the result of a galaxy's interaction with its
environment (Dressler 1980). For example, galaxy mergers are believed 
to transform spiral galaxies into ellipticals (Tommre \& Tommre 1972, 
Barnes 1990) 
and the removal of gas from galaxies is thought to drive the 
transformation of spiral galaxies into the S0 types (Gunn \& Gott, 1972,
Larson et al.\ 1980).

The ``inverse'' of such transformations
is less widely considered. Yet, the
regeneration of spiral disks in elliptical galaxies is a key component of 
current galaxy formation models (White \& Frenk 1991, Cole et al.\ 1994, 
Kauffmann et al.\ 1993, Sommerville \& Primack, 1999, De Lucia et al.\ 2006, 
Bower et al.\ 2006). In such models, galaxies vary in morphology as they grow 
potentially passing through several cycles back and forth across
the Hubble diagram. This paper is the first in a series examining the 
observational evidence of such morphological cycles. Here, we focus on the 
gas-rich dust-lane elliptical NGC~3108, asking whether this could be an 
example of a ``hermaphrodite''\footnote{In biology, a hermaphrodite is an organism that possesses both male and female genitalia during its life.  Some are able to change sex multiple times.  {\it Sequential hermaphrodites} are organisms born as one sex and then later change into the other sex, an example being the Clownfish (source: www.wikipedia.org).  We find hermaphrodite a useful analogy to galaxies undergoing multiple transformations in a hierarchical Universe. We decided to use this term because our criteria go beyond the observationally motivated labels; for examaple, ``disky'' and ``S0'' do not describe the direction of the transformation.} galaxy under-going such an ``inverse'' morphological 
transformation.

Of course, on-going star formation in elliptical galaxies is not a new
discovery. Many ellipticals have small amounts of ionized gas and star 
formation (eg., Caldwell 1984, Knapp et al.\ 1985, Philips et al.\ 1986, 
Goudfrooji et al.\ 1994, Sadler et al.\ 2000, Buyle et al.\ 2006). 
However, the star formation is usually low-level and centrally concentrated, 
or associated with features indicating a recent merger event: of the 
galaxies in the SAURON HI 
survey with significant ionized gas content, only the polar ring galaxy
NGC~2685 has an HI gas mass greater than $4\times 10^7$ \msun 
(Morganti et al., 2006). Low levels of on-going star formation
 are very likely to be driven by the 
reprocessing of stellar mass-loss during the late phases of stellar 
evolution, or the accretion of small gas rich companions (Emsellem et al.\ 2007). 
Recent observations from GALEX seem to confirm this view.  
The large scatter in the near-ultraviolet colour-magnitude relation suggests
that roughly 15\% of $z < 0.13$ bright ($M(r) < -22$) early-type galaxies
show signs of recent ($\ale$ 1 Gyr) star formation at 1--2\% level in
mass compared to the total stellar mass, which implies that recent
low-level residual star formation may be common even in bright early-type
galaxies (Yi et al.\  2005, Salim et al.\ 2005). Such ``frosting''
is also supported by 
other studies (e.g. Trager et al 2000; Terlevich \& Forbes 2002; Koo et al 2005) and the existence of 
young stellar disks at the centres of many elliptical galaxies is also 
becoming well established (Scorza et al 1998, Hau, Carter \& Balcells 1999, Emsellem et al.\ 2007). 

However, while such residual activity is relatively common, these 
small amounts of star formation generate far fewer stars than the existing 
bulge and will never result in a significant change in the 
galaxy's morphology or bulge-to-disk ratio.
In order to result in a genuine ``inverse'' transformation, the new
episode of star formation must have an extent greater than the 
existing spheroid, and must eventually generate sufficient new stars to 
significantly lower the relative importance of the bulge light. Such galaxies are likely to be much more 
common in the distant universe, where the rates of merging and gas
accretion are much higher than today. However, the concept of such cycles 
of morphology will receive considerable support if even one {\it bona fide} 
example of an ``inverse'' morphological transformation could be established
in the local universe.

Dust-lane ellipticals may be a good starting point for this search.
Osterloo has observed such galaxies, discovering that several contain
substantial HI gas mass distributed in a macroscopic disk 
(Oosterloo et al.\ 2002, 2007, J\'ozsa et al.\ 2004). One prominent
example is NGC~3108. In this paper we report on deep optical observations
of this galaxy, probing the star formation associated with the 
HI gas disk.  In another paper, we will report the results of parallel
search in a sample of isolated elliptical galaxies.

The structure of this paper is as follows. We summarise the available
data on NGC~3108 in \S2 and present our H$\alpha$ imaging in \S3. We
derive star formation rates in \S4, and consider the HII region luminosity
function and the age of the stellar disk in \S5 and~6. In the discussion 
(\S7) we examine whether NGC~3108 is a plausible candidate for a 
hermaphrodite galaxy and compare it with two other nearby candidates NGC~5128
and the Sombrero galaxy (M104). 
We also briefly consider the implications in the context of current galaxy 
formation models. The conclusions are summarised in \S8.

\section{NGC 3108}

NGC 3108 was chosen for H$\alpha$ imaging because of its high-HI
content (Oosterloo et al 2002; hereafter O02). It resides in a small galaxy group LGG187 (Garcia 1993) where four, including NGC 3108, were detected in HI. The other
detected galaxies are two spirals and a low-surface brigtness, hydrogen-rich
galaxy. Although classified as S0p (de Vaucouleurs et al 1991), its light profile follows closely that of a $r^{1/4}$ law (Caldwell 1984; see also \S~\ref{sec:mass}). NGC 3108 has a minor-axis dust lane, which is unusual  since an infrared image from 2MASS only shows a bulge of P.A. $\sim 45\deg$, and bulges are usually aligned with the disk major axis.  O02 remarked that
two dust lanes can be seen in optical image, along with a possible
shell. Caldwell (1984) also detected a disk of ionized gas in the
inner 10\asec. NGC 3108 possesses a large amount of HI ($2.3 \times
10^9 $ \msun; $H_o = 72$ \kms Mpc$^{-1}$), with the neutral
hydrogen quite extended and regularly distributed and orientated
perpendicular to the optical major axis. Its $M_{{\rm HI}} / L_{\rm B} $ ratio is high compared to old merger remnants and ellipticals, and more similar to the values for very young ($ < 1$ Gyr) merger remnants (Georgakakis et al 2001).
 The HI is distributed in a
regular, rotating disk-like structure extending $\sim 20$ kpc in radius, with a central hole, and may be warped in the outer parts. The HI has
a wide ($\pm 300\ \kms$) velocity range.  The average surface density of HI is estimated by O02 to be about $0.5$ -- $1\
\msun\ pc^{-2}$, and O02 concluded that it is too low for large
scale star formation, maybe with the exception of localised regions.
 The lack of radio continuum emission with $3
 \sigma$ upper limit of $\sim 1.4$ mJy at 1.4 GHz led O02
 to set an upper limit to the SFR of 1 \msun yr $^{-1}$.

\section{H$\alpha$ Observations}

\label{sec:halpha}

During 2005 December, H$\alpha$ and R-band images of NGC
3108 were taken with the EMMI instrument on ESO's NTT at the La Silla Observatory.
The H$\alpha$ 597 filter is centred on the redshifted H$\alpha$ line
at 3120 \kms, with a FWHM of 3020 \kms. With each filter two deep exposures were taken
 with a telescope offset for cosmic ray removal. A shallow exposure is taken to replace the saturated
pixels. The total exposure times for H$\alpha$
and R are 2060 and 360 seconds respectively. The night was
photometric, and the spectrophotometric standard star Hiltner 600 was
imaged immediately after the galaxy exposure for flux calibration. The seeing was 0.8\asec.

The data were reduced following standard procedures. 
Bad and saturated pixels
are tracked throughout by a bad pixel mask. 
Regions near bright sources are also masked out to avoid light
contamination. Special attention was paid to sky subtraction and any residual
background gradient after flat-fielding was removed using a low-order
polynomial fit to the sky away from the galaxy. The fluxes were corrected for atmospheric extinction (using the typical R-band
extinction coefficient, $k_e = 0.1$ for EMMI) and interstellar
extinction (using estimates from Schlegel, Finkbeiner \& Davis
(1998)). The fluxes were finally converted to units of
e$^-$ s$^{-1}$ pixel$^{-1}$.

\begin{figure*}
\includegraphics[scale=0.75]{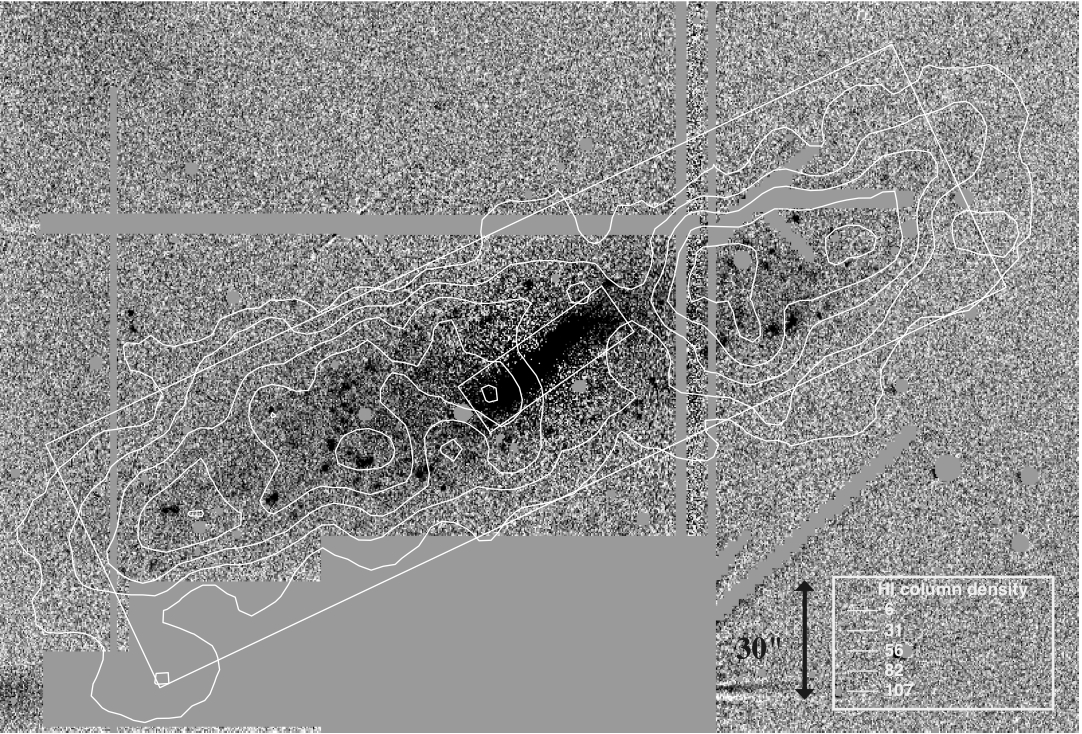}
\caption{ Pure H$\alpha$ emission map of NGC 3108 over laid with contours
of HI gas density. The grey areas represent masked regions excluded from the
emission analysis. The 2 rectangles define the areas where the flux is
measured for the inner bar and the entire disk. The white contours
correspond to HI column densities in units of $10^{19}$ atoms
cm$^{-2}$. North is up and East is left. 30\asec corresponds to 5.38 kpc.
}
\label{figure:N3108Emis}
\end{figure*}

\begin{figure*}
\includegraphics[scale=0.75]{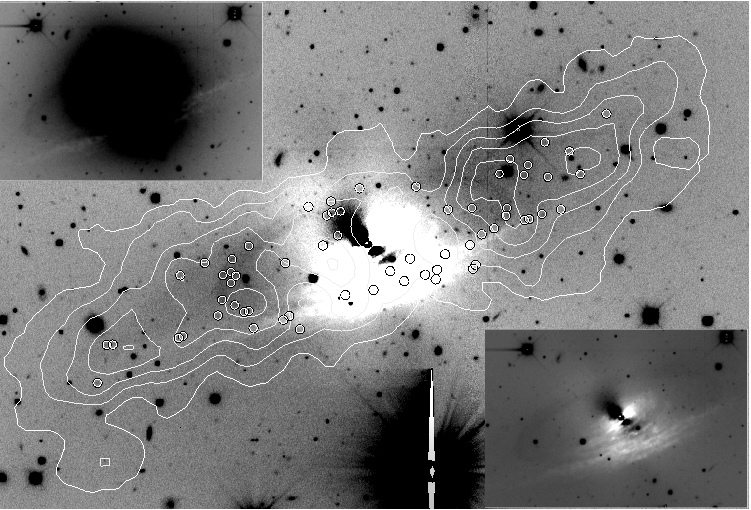}
\caption{ Main image: R-band image of NGC 3108 with a Sersic model
subtracted, showing the faint stellar disk and the dust
structure. Black is positive and white is negative. Overplotted in
white are HI contours, and the HII regions detected in
\S~\ref{sec:HII} are marked by circles. Top insert: R
band image of the galaxy before subtraction. 
Bottom insert: the same as the main image
but with cuts adjusted to reveal the inner dust structure. Both
inserts measure $134\asec \times 90\asec$.}
\label{figure:N3108R}
\end{figure*}

The procedures for obtaining the pure emission images from the narrow
and R-band images follow those in De Rijcke et al (2003; hereafter
DZDH03), with the following exception. As with DZDH03, the pure
emission $Em$ is recovered from a narrow-band image $Nb$ and an R-band
image $Rb$ as
\begin{equation}
Em= Nb - (c  \times Rb + \delta)
\end{equation}
where $c$ is a scaling constant and $\delta$ is a correction for
 imperfect sky-subtraction. DZDH03 fit isophotes to the narrow-band
 and the R-band images and determine the optimal values for $c$ and
 $\delta$ by minimising the $\chi^2$ difference between the
 isophotes. This is possible as DZDH03's sample consist of dwarf
 ellipticals with little star formation, thus the isophotes for the 2
 filters are very similar. NGC3108 has extended H$\alpha$ emission and
 DZDH's method will result in too much of H$\alpha$ being subtracted off. After
 extensive experimentation we find the following works best. 
 $\delta$ is set to zero as sky subtraction was good. $c$ is allowed to vary and the resulting emission map is examined.
 As the major axis of the galaxy is along the disk axis, too much bulge light will be subtracted off if $c$ is overestimated, and the resulting image will have 2
 negative ``lobes" perpendicular to the disk. If $c$ is underestimated,
 then the 2 ``lobes'' perpendicular to the disk become positive. We find
 with this method $c$ can be found to within about 1\%. This accuracy
 is more than enough considering the other uncertainties in the flux
 calculation.
 As a sanity check,
 the value of $c$ adopted is very close to those calculated for other
 elliptical galaxies at similar redshifts observed with the same
 filter, results of which will be presented in a
 future paper.

The pure emission image is shown in Fig.~\ref{figure:N3108Emis}. It is evident that the
emission consists of 2 components. In the inner regions the emission is
in the form of a bar of 43\asec (7.7 kpc) in extent. This coincides with the ionized gas ``disk" that Caldwell (1984) found from long-slit spectroscopy along P.A. = 120\deg. On much larger
scale, there is a faint disk along the minor axis, with a maximum extent of 220\asec (39 kpc). The inclination of the HI disk axis with respect to the line-of-sight is about 73\deg. Superimposed on the disk
are a number of HII regions.

The spatial distribution of the HII regions bears good resemblance with a
barred spiral. Two spiral arms can be traced: the eastern arm emerges
from the NW tip of the bar, goes over the north side of the
nucleus, towards the SE, then S and finally W. The western arm
emerges from the SE tip of the bar and goes towards the
NW. The spiral arm structure is even more striking when Fig.~\ref{figure:N3108Emis} is compared with Fig.~\ref{figure:N3108R}.
 Figure~\ref{figure:N3108R} shows that the detected HII regions
populate the edges of the dust lane, along roughly a brightened ring around the galaxy. They resemble spiral arms
where young star clusters populate near the edges of dust lanes. The bar is heavily obscured by dust.

In Fig.~\ref{figure:N3108Emis} HI column density contours from J\'{o}zsa et al 2004 are overplotted. It is
remarkable that the large outer emission disk coincides with the HI
emission. Furthermore, the hole in the centre of the HI emission
corresponds to the location of the bar, although
the bar does not coincide with the centre of the hole, but offset SE
from it. This could be a projection effect, with the bar in the foreground of the eastern spiral arm.
 On large scales, there is a good agreement between the spatial distribution
of the H$\alpha$ emission and the HI. The locations of maximum HI column density correspond to the tangential points of the spiral arms; this seems to suggest that the HI traces the spiral arms. 

\section{Star formation rates}

The CCD signal is converted to H$\alpha$ flux following the procedures of DZDH03.
 Since the observed emission flux is a composite of the H$\alpha$ 6563
\ang, [NII] 6548 \ang\ and [NII] 6583 \ang\ (whose flux will be denoted by
 $F_{{\rm H}\alpha}$, $F_{{\rm [NII]}_1}$ and $F_{{\rm [NII]}_2}$ respectively), the
 uncertainties for converting the CCD signal to $F_{{\rm H}\alpha}$ depend
 on the ratios $F_{{\rm [NII]}_2}/F_{{\rm [NII]}_1}$ and
 $F_{{\rm [NII]}_2}/F_{{\rm H}\alpha}$. The former is set to 3 (Macchetto et al
 1996; Phillips et al 1986, DZDH03). 
The latter is set to 0.54, measured directly from a 8100 seconds long-slit spectrum of an HII region of NGC 3108 taken with the EFOSC2 instrument on the ESO 3.6 telescope in January 1999.  

The total $F_{{\rm H}\alpha}$ is measured separately, for the ``inner bar"
and ``entire disk", defined by the 2 rectangles in Figure~\ref{figure:N3108Emis}. We
obtain:
\begin{enumerate}
\item[]  $F_{{\rm H}\alpha} ({\rm Inner\ bar}) =  5.76 \times 10^{-14}$ \ergsqcmsec
\item[]  $F_{{\rm H}\alpha} ({\rm Entire\ disk}) = 1.74 \times  10^{-13}$ \ergsqcmsec
\end{enumerate}
To covert from $F_{{\rm H}\alpha}$ to total luminosity we adopt a
Hubble Flow distance (corrected for Virgo, Great Attractor and Shapley infall) from NED\footnote{The NASA/IPAC Extragalactic Database (NED) is operated by the Jet Propulsion Laboratory, California Institute of Technology, under contract with the National Aeronautics and Space Administration.}  of
 $39.0$  Mpc, and obtain:
\begin{enumerate}
\item[]  $L_{{\rm H}\alpha} ({\rm Inner\ bar}) =  1.05 \times 10^{40}$ \ergsec
\item[]  $L_{{\rm H}\alpha} ({\rm Entire\ disk}) = 3.15 \times  10^{40}$ \ergsec
\end{enumerate}

Finally we use Kennicutt's relation (Kennicutt 1983) to derive the total Star Formation Rate (SFR):
\begin{equation}
SFR = \frac{L_{{\rm H}\alpha} A_{{\rm H}\alpha}}{1.12\times10^{41}} \msunyr 
\end{equation}
where $A_{{\rm H}\alpha}$ is the unknown internal extinction factor.  We use
the empirical B-band attenuation-inclination relation for disks
derived from the Millennium Galaxy
Catalogue (Driver et al 2007):
\begin{equation}
\Delta A_B = 0.99 \times [1 - {\rm cos}(i)]^{2.32} + A_{{\rm B, face-on}}   \label{eqn:driver}
\end{equation}
where $i$ is the inclination of the disk ($0=$ face-on), and $A_{{\rm B, face-on}}=0.2$ is the face-on attenuation for disks. Taking $i=73\deg$ (\S~\ref{sec:halpha}), $A_{\rm R}/A_{\rm B}$= 0.64 (Knapen et al 1991), and by assuming $A_{H\alpha} = A_R$, $A_R$ is estimated to be  0.41 mag or a factor of 1.5. 
This compares very well with $A_R = 0.42$ calculated from Caldwell's (1984) estimation of the average $A_V$ for the dust lane.  
We get:
\begin{eqnarray}
SFR ({\rm Inner\ bar})   =  0.14\ \msunyr  \\
SFR ({\rm Entire\ disk})  =  0.42\ \msunyr  \label{eqn:totsfr1}
\end{eqnarray}
Thus the outer disk is forming stars at twice the rate as the inner
bar. 
The total SFR is comparable to that of the H$\alpha$ derived SFR of M81
($0.38$ \msunyr; Gordon et al 2004), a nearby Sab galaxy, and also consistent with the upper limit set by O02. 

The extinction value $A_R=0.42$ adopted above is an estimation for the averaged extinction over the galaxy. It may be possible that the extinction is higher for the inner bar and the HII regions where the H$\alpha$ flux originates, such that $A_{H\alpha}$ is higher than $A_R$. We note that as $SFR  \propto A_{{\rm H}\alpha} $, a doubling of the extinction will double the $SFR$.

\begin{figure}
\includegraphics[scale=0.35]{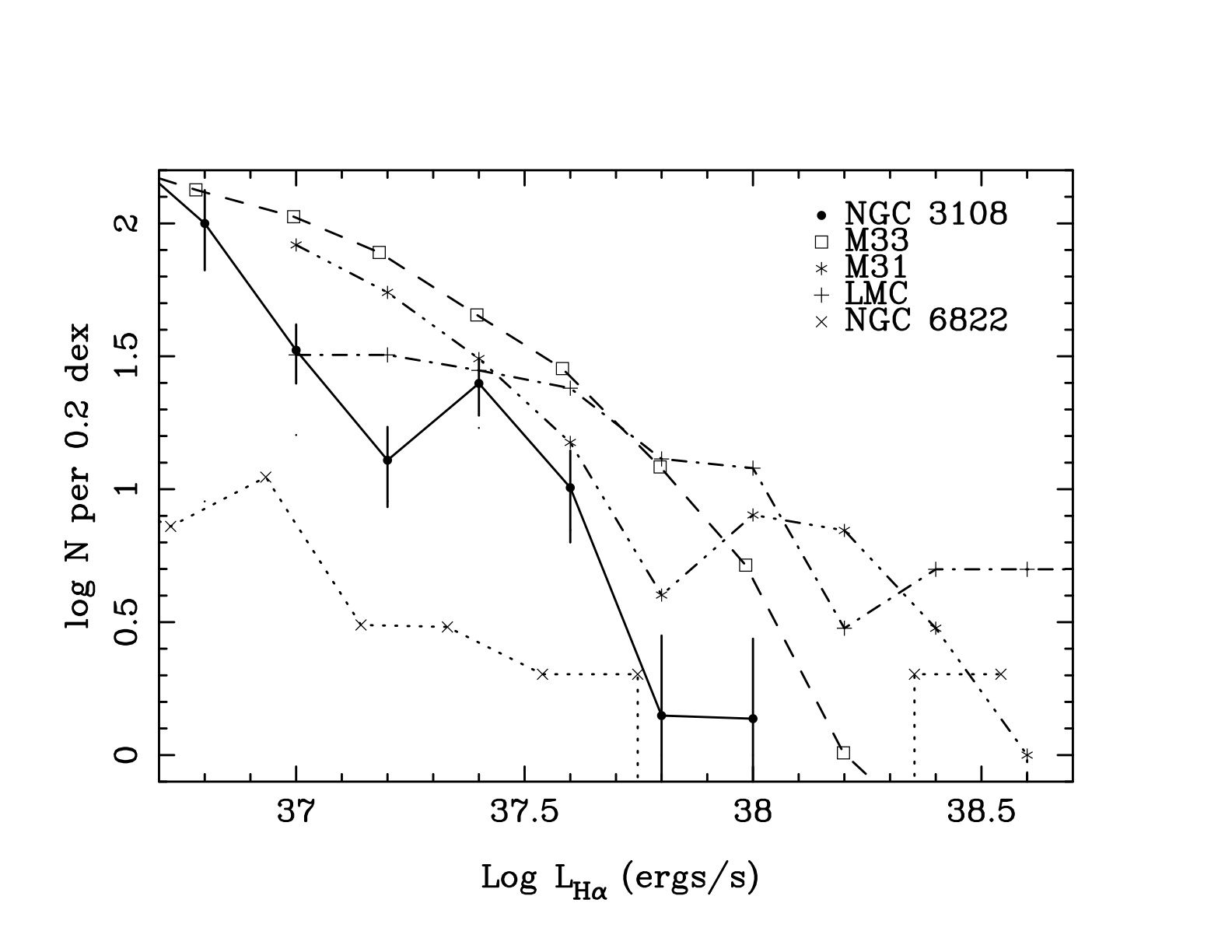}
\caption{ The HII regions  luminosity function of NGC 3108, corrected for incompleteness (large
dots and solid line), with a binning of 0.2 dex. Plotted are also the
HIILF of other galaxies in the Local Group: M31 \& LMC (Kennicutt,
Edgar \& Hodge 1989), M33 (Hodge et al 1999), and NGC 6822 (Hodge, Lee
\& Kennicutt 1989). The luminosities are corrected for Galactic
extinction but not for internal extinction of each galaxy.  }
\label{figure:N3108HIILF}
\end{figure}

Kennicutt (1989) suggested that star formation could not occur in systems
of very low gas surface density. It is interesting to compare the gas
disk in NGC~3108 with the proposed threshold of  3--4 \msun pc$^{-2}$.
Assuming the $2.3\times10^9$  \msun\ of HI is in a uniform circular disk, 
the surface density is 2 \msun pc$^{-2}$. However, the HI is unlikely to 
be distributed uniformly and will be concentrated in spiral arms where 
the local surface density may well exceed Kennicutt's threshold. We note from 
Figure~\ref{figure:N3108Emis} that the
H$\alpha$ emission is confined to within HI contours of $31 \times 10^{19}$ 
atoms cm$^{-2}$, which translate to a projected an averaged surface 
density of 2.5  \msun pc$^{-2}$, which is still below Kennicutt's threshold. We conclude that the star formation
in NGC 3108 is occuring in a unusually low gas density disk, however, the
star formation may trace regions with slight enhancements of the gas density
where the density threshold criterion is met.

\section{HII regions}\label{sec:HII}

Because of the uniqueness of this system, it is interesting to
investigate the size and luminosity distribution of individual HII
regions. {\tt Sextractor} was run to extract sources with significance
level above 2.5 $\sigma$. The detected HII regions are plotted in
Figure~\ref{figure:N3108R}.  Detection incompleteness is
estimated by Monte Carlo simulations using artificial sources of
uniform spatial distribution, and are 73\% and 48\% at
log $L_{H\alpha} = 38$ and 37 dex respectively. The resulting HII regions
Luminosity Function (HIILF), uncorrected for any internal extinction, is plotted in
Figure~\ref{figure:N3108HIILF}, and is compared with the HIILF's of
Local Group galaxies. The HIILF of NGC 3108 most resembles that of the
Sc galaxy M33 in shape, with a lack of bright HII regions with log
$L_{{\rm H}\alpha} > 38$ dex excited by large star clusters, unlike M31 and LMC. NGC
3108's HIILF will agree with that of M33 if the HII regions are about
3 times more numerous, or if the internal extinction in NGC 3108 is
0.5 mag higher than that in M33, which is seen more face-on. Note that the expected difference in $A_R$ between disks with $i=0$ and $i=73\deg$ is 0.28 mag, using equation \ref{eqn:driver}.

We fit 
the power
law $N(L) = A L^{\alpha} dL$ to points with log $L_{{\rm H}\alpha}  \geq  37$ dex, and obtain $\alpha = -2.1$.
This is within the range spanned by spirals and irregular galaxies
(Kennicutt, Edgar \& Hodge 1989).

The HII regions in NGC 3108 are unresolved. This sets an upper limit
of 150 pc to their sizes. In comparison the 3 brightest HII regions in
the Local Group (30 Doradus, NGC 604 \& NGC 395) have sizes
$\approx100$--$200$ pc (Wilson \& Scoville 1992; Kennicutt 1988;
Wilson \& Matthews 1995).  The properties of the HII regions in NGC 3108 are consistent with those in disk galaxies. The mode of star formation seems normal.

\section{Mass and age of the existing stellar disk}\label{sec:mass}

The image in the top inset of Figure~\ref{figure:N3108R} shows that the major axis of the galaxy
lies roughly along the NE direction. However at
very faint levels there is excess light along the direction of the dust
lane (cf., Caldwell 1984), and it is well matched in spatial distribution with 
the HI and H$\alpha$ emission. This suggests that a significant population 
of stars have already formed out of the gas and that the star formation 
discussed above is part of a continuous process.

The faintness of the stellar disk compared with the bulge, and the extensive dust features prevents any meaningful multi-component photometric decomposition. Instead,
the luminosity of the stellar disk is estimated by fitting and
subtracting a Sersic model from the R-band image using the program
{\tt GALFIT} (Peng et al 2002). The best fit model (the ``bulge'')  has a Sersic index $n$
of 3.64, $R_e$ of 23.7\asec, axial ratio of 0.86, P.A. of
48.5\deg, and apparent magnitude of 11.2 mag. The residual
image is shown in Figure~\ref{figure:N3108R}. We use this to measure the disk
surface brightness away from the galaxy centre (to avoid regions where 
some of the bulge light may have been over-subtracted), and away from the 
dust features. 
The surface brightness is then scaled with the disk area, assuming a constant projected surface density, to obtain a total magnitude
of $15.1 \pm 0.3$ mag. The corresponding total luminosities for the
bulge and the stellar disk
are $L_{\rm R} =1.4 \times 10^{11}$ and $2.1 \times 10^9$ \lsun\
respectively if the R-band internal extinction for the bulge and the disk are
1.1 and 0.4 mag respectively, derived using the empirical relations
from Driver et al (2007).
 To convert
from luminosity to mass, we use $M/L$ ratios of single stellar
populations with solar metallicities from Worthey (1994). The bulge
stars are assumed to be 10 Gyr old, and the corresponding bulge
mass is $3.6\times 10^{11}$ \msun, corrected for internal
extinction. If the stars are 6 Gyr old, the corresponding disk stellar mass is $3.9
\pm 1.2 \times 10^9$ \msun. If the stars are 1 Gyr old, the disk mass is $1.4 \times 10^9$ \msun.
The former compares well with the mass estimated for the ``luminous ring"  by Caldwell (1984), which would be $3.1 \times 10^9$ \msun\ if corrected for the difference in adopted distance moduli, Galactic foreground and internal extinctions, and the fact that Caldwell used $M/L=1$ and the V band. 
Thus the current disk-to-bulge mass ratio
is $\sim 0.01$.

A rough star formation timescale of the stellar disk can be estimated by
assuming that all the disk stars have been formed from HI with the global, constant SFR
given by  equation~\ref{eqn:totsfr1}, as the total mass divided by the
SFR. Assuming the extinction in the stellar disk in  R band is equal to that in
H$\alpha$, the age of the disk is $9.4\pm3.1$ Gyr. 

A better way to estimate the star formation timescale is to measure  the {\it local} surface brightness of the stellar disk in R and the corresponding H$\alpha$ flux within the same
window, as long as the H$\alpha$ is well sampled. 
The local disk luminosity
is converted to mass assuming a R band $M/L$ ratio and then
divided by the locally derived SFR. Furthermore we assume that the
extinction is the same in stellar disk and the HII regions.  
The star formation timescale derived this way is much larger than the stellar
age put into the model: for a 6 Gyr old, solar metallicity population,
the timescale is $27\pm9.4$ Gyr. To bring it down to 6 Gyr, the SFR has to be boosted by assuming that the H$\alpha$ extinction is higher than that in R by 1.6 mag, presumably because the internal extinction within HII regions is higher than that averaged over the disk.
  Such high relative extinction is
unlikely since this would boost the global SFR above the 1 \msunyr\  limit 
set by the the 1.4 GHz observations
(O02). While it is possible that the H$\alpha$ flux may suffer an extra extinction, the large age discrepancies suggest that the existing stellar disk is too massive to have been formed by the current SFR. We conclude that the SFR must be declining, and that the disk SFR was higher in the past.

\section{Discussion}

\subsection{NGC 3108: a  failing ``hermaphrodite''}

We have seen that NGC 3108 contains a substantial gas disk. In theoretical models this might come from gas accreted from a hot or rapidly cooling halo (White \& Frenk 1991), or cold gas accreted from filaments (Kere\v{s} et al 2005). Other possible origins might be gas left over from the merger of gas-rich progenitors or accretion of gas rich dwarf satellites (Navarro 2004), although the amount of HI in NGC 3108 is probably too much for a single dwarf. The key issue is not really the origin of the gas, but rather whether the rate of gas supply is sufficient to grow a massive disk and turn NGC 3108 into a spiral morphology. 

How long can NGC 3108 continue to grow a disk? Its HI mass is $2.3
\times 10^9$ \msun, so it can continue to form stars at current rate
for about 5.5 Gyr without recycling or accretion. Will NGC 3108 turn back into something resembling a spiral galaxy by the time the current HI gas reservoir is used up? The
stellar disk currently has a mass of $3.9\times 10^9$ \msun\ and it will
grow by 60\% to $6.2 \times 10^9$ \msun, about $1/5$ of that of
the Milky Way ($\sim 3\times 10^{10}$ \msun\ assuming a 3 kpc
scalelength; Flynn et al 2006). While this is substantial, it is still
insignificant compared with the large bulge mass: the final
disk-to-bulge stellar mass ratio would be 0.02. 

In order to make a firm morphological transition the star formation rate and
gas accretion rate would need to be increasing. This appears to be ruled
out. Our comparison with the disk continuum showed that the disk was old
and that the star formation rate was declining. This evidence suggests that
the accretion event that created the new disk was a one-off occurrence, 
possibly even associated with the formation of the bulge. A plausible 
scenario is that the gas is residual high angular momentum material that
was spun out of the system during the merging event 
(Mihos \& Hernquist 1996; Barnes 2002; Robertson et al.\ 2006).

What does the future hold for NGC 3108? It will remain a
dust-lane S0 with a weak stellar disk, unless the SFR can be boosted and there is more gas accretion in the future. However, the SFR seems to be declining, suggesting that additional gas accretion (which should boost the SFR) has not been significant at least up to the present. In deep images, NGC 3108 will be 
classified as a S0 or disky elliptical/fast rotator (Emsellem et al 2007).

\subsection{Comparison with two nearby ``hermaphrodite" candidates}  \label{sec:compare}

It is interesting to compare NGC~3108 with two other 
hermaphrodite candidates: NGC 5128 (Centaurus A) and M104 (Sombrero).  These
systems also show evidence that could be interpreted as a macroscopic disk
growing around a bulge system. 

Morphologically, NGC 3108 and NGC 5128 bear some resemblance, with
prominent dust lanes.  NGC 5128's dust emission looks like a barred
spiral residing inside an elliptical (Mirabel et al 1999).  The size
of the inner bar of NGC 5128 (1 kpc) is smaller than that in NGC 3108
(3.9 kpc), and in both galaxies the outer ends of the bar are
connected to trailing spiral arms, defined by the large scale dust
lanes. However, NGC 5128 has prominent shell structures and an arc of
young blue stars (Peng et al 2004), indicative of recent merger
activity. Such features are lacking in NGC 3108.

The Sombrero galaxy also has a strong inner bar (within the inner
Linblad Resonance; Emsellem \& Ferruit 2000), but the outer spiral
arms are tightly wound (van der Burg \& Shane 1986) so much so that
they look like 2 partially overlapping rings (Bajaja et al 1984).
Sombrero also has a much more massive stellar disk ($5.5 \times
10^{10}$ \msun\ corrected to $H_o =72$ \kms Mpc$^{-1}$; from van der
Burg \& Shane 1986), which may have at least partly contributed to the
bulge through secular evolution driven by a now dissolved stellar bar
(Emsellem 1995; Emsellem et al 1996). Despite the mass of the stellar disk,
the Sombrero has a high bulge to total ratio, with estimates ranging from 
0.73 (Baggett et al 1998) to 0.85 (Kent 1988).

With a HI mass of $4.5 \times 10^8$ \msun\ (van Gorkom et al 1990;
Schminovich et al 1994), NGC 5128 does not have as much HI as NGC
3108. Meanwhile Sombrero has $1.4 \times 10^9$ \msun\ (converting van
der Burg \& Shane (1986)'s value to correspond to a distance of 19.4
Mpc), which is comparable to that of NGC 3108. Despite its low gas mass, 
NGC 5128 has a much higher SFR than NGC 3108 (at 2 \msunyr\ ; Telsco 1978;
Tinsley 1980), while that of the Sombrero is about the same (at 0.41
\msunyr\ using Hameed \& Devereux (2005)'s (H$\alpha$ + NII) luminosity
and correcting for extinction using Driver et al (2007)'s formula; and
at $\sim 0.8$ \msunyr\ estimated from 1.49 GHz observation (Bajaja et
al 1988)). As with NGC~3108, both NGC 5128 and the Sombrero will soon
run out of HI gas unless fresh material is added to the system. At
current SFR, NGC 5128 will exhaust its HI gas reservoir in 0.2 Gyr
while the Sombrero will do so within the next 3 Gyr, while increasing
its disk mass only by another 2\%.  At the current star formation rate
the Sombrero would take 145~Gyr to build the stellar disk: the current star formation rate in
the Sombrero must be at least a factor 10 lower than the peak disk 
star formation rate.

The lack of a substantial HI reservoir, the higher current SFR (than
that in NGC 3108), and the lack of a prominent stellar disk suggests
that NGC 5128, despite the good morphological resemblance, is not a
good hermaphrodite candidate. Its current star formation is likely to be
associated with recent merger activities possibly due to recent
accretion of a gas rich dwarf galaxy (Peng et al 2004) or the
interaction of the radio jet with the gas clouds deposited during the
last merger (Rejkuba et al 2002). Its current energetic activities
(Isreal 1998; Kraft et al 2004), a much smaller gas reservoir, and a
more compact bar structure bear more resemblance with short-lived
merger-induced star formation than prolonged, orderly star formation
capable of generating a substantial stellar disk. The barred spiral in
NGC 5128 is probably a quasi-stable system formed from tidal debris of
gas-rich object(s) accreted in the past $10^9$ years, and currently
forming a ``symbiotic" relationship with the host elliptical (Mirabel
et al 1999).  On the other hand, the Sombrero galaxy, with its 
substantial stellar disk, similar SFR to NGC 3108, comparable HI 
reservoir and lack of evidence for recent merger activities, resembles more closely 
NGC~3108. However, the disk star 
formation rate appears to be declining rapidly and further disk growth has
stalled. It is tempting to suggest 
that it represents a more advanced stage of evolution of a ``failing'' 
hermaphrodite like NGC 3108.

\subsubsection{Implications for galaxy formation models}

The existence of ``hermaphrodite'' galaxies which are in the process of 
transforming from early to late type morphology
is a key but untested prediction of hierarchical galaxy formation
models (e.g. White \& Frenk 1991; Kauffmann, White \& Guiderdoni 1993;
Cole et al 1994; Sommerville \& Primack 1999; Bower et al 2006; Croton
et al 2006). The widely successful semi-analytic models combine a 
set of simple rules describing the gas processes involved in galaxy 
formation, with a scheme to follow the hierarchical growth of dark matter 
haloes. To form an early-type galaxy, the models assume that
gas initially cools into a disk from which stars form. When two dark
matter haloes merge, the galaxies within them can coalesce. If the
dynamical friction timescale for a galaxy is shorter than the lifetime
of the dark matter halo it resides in, then the galaxy merges with the
central galaxy in the halo. During a violent merger, the disk of the
central galaxy is destroyed and the stars are moved to the bulge, and
any cold gas present is turned into stars in a burst and added to the
bulge. This is the conventional sense of morphological
transformation--- forming ellipticals from merging spirals is an old
idea (Toomre \& Toomre 1972). Immediately after a violent merger, the
galaxy will have a pure bulge morphology. However, depending on the subsequent
evolution of the halo, more cool gas may be accreted in a disk and
turn into stars, increasing the disk-to-bulge ratio of the
galaxy. Thus, over a Hubble time, the galaxy may turn back into a
system with low bulge to total mass ratio.
 
This process is generic to semi-analytic models, but it is yet
to be observationally established. One problem is that much of the
action may have occurred at high redshifts when star formation and 
gas accretion rates are higher. Nevertheless it is
puzzling that so few hermaphrodite candidates are seen in the local
Universe. 
Our observations of NGC~3108 fail to establish it as a example 
of such a galaxy. Likewise, the Sombrero galaxy and NGC5128 (Centaurus A) 
fail to meet the criteria we have laid out. While it is beyond the scope 
of this paper to make a detailed comparison with the space densities 
predicted by the models, it is useful to qualitatively consider the 
implications of a null result. Would this invalidate the successes of
the hierarchical model?  There are two important caveats: (1) the gas 
accretion phase may occur very rapidly after the galaxy merger making it
hard to distinguish ``fresh'' gas which is capable of growing a 
new disk from material that is left over from the merger event;
(2) spiral bulges may be primarily formed by secular disk 
instabilities (Toomre 1964; Efstathiou, Lake \& Negroponte 1982) 
so that the outer disk is never completely disrupted. (In this scenario, the 
mergers that generate pure bulge systems may dominate only in environments 
where further star formation and gas accretion is strongly 
suppressed, Bower et al.\ 2006). Clearly it is important to carefully take 
these factors into account when comparing with the theoretical models.
We intend to make this quantitative comparison in a following paper based
on a systematic survey of isolated elliptical galaxies.

\section{Conclusions}

We investigated whether NGC 3108, one of the more promising candidates for a ``hermaphrodite'', is an early type galaxy which has been slowly transforming into a spiral galaxy. We have argued that such ``inverse morphological transformations" are a fundamental component of galaxy formation models but are yet to be proven observationally.  Using H$\alpha$ and R band imaging, we derived an extinction-corrected SFR of 0.42 \msunyr. The emission is in two components: a nuclear bar which coincides with the hole (and offset from it) in the HI distribution, and an extended disk component which coincides with the HI distribution and mainly confined to HI densities of $31 \times 10^{19}$ atoms cm$^{-2}$. This suggests that a stellar disk is currently being formed out of the HI gas. The HII regions are located close to the dust lanes and define a spiral structure; comparison with the HI distribution suggests that the HI is also distributed in a spiral arm structure. The mode of star formation is similar to that in other spirals, with the HII regions luminosity function most resembling that of M33 in shape.

We measured the current disk mass and find that it is probably too massive to be formed by the current star formation rate over the last few Gyr. While we cannot rule out that the current star formation may be more obscured than estimated, the amount of extinction required is too large. More likely that the SFR in NGC 3108 was higher in the past. With the current SFR, the  disk in  NGC 3108 will grow to be $6.2\times 10^9$ \msun\ in stellar mass within a Hubble time. While this is substantial, the disk will be insignificant compared with the large bulge mass: the final stellar mass disk-to-bulge ratio will be $\sim 0.02$. NGC 3108 will fail to transform into anything resembling a spiral without a boost in SFR and additional supply of gas.

NGC 3108 is compared with two other ``hermaphrodite'' candidates. Although there is a good morphological resemblance with NGC 5128, a ``symbiotic'' galaxy resembling a barred spiral residing inside an elliptical, the latter is more typical of a merger remnant with a short period of intense activities.
The Sombrero galaxy with its dissolved outer bar and more massive stellar disk may represent the end point of a ``failed hermaphrodite''.
All 3 galaxies will be unable to grow a sufficiently large disk to lower
their bulge-to-total ratio unless there is a fresh supply of HI gas and the 
SFR can be boosted. In contrast, the observations suggest that the current
episode of star formation will be short lived and ineffective.

\section{Acknowledgements}

We thank Marina Rejkuba and Gretchen
Harris for useful feedback and suggestions. DAF thanks the Australian
Research Council for financial support. MLB is supported by a Discovery
Grant from the National Science and Engineering Research Council
(NSERC).  This research has made use of
the NASA/IPAC Extragalactic Database (NED) which is operated by the Jet
Propulsion Laboratory, California Institute of Technology, under
contract with the National Aeronautics and Space Administration.


\begin{thebibliography}{}

\bibitem[\protect\citeauthoryear{Baggett, Baggett, \& 
Anderson}{1998}]{1998AJ....116.1626B} Baggett W.~E., Baggett S.~M., 
Anderson K.~S.~J., 1998, AJ, 116, 1626 

\bibitem[\protect\citeauthoryear{Bajaja et al.}{1984}]{1984A&A...141..309B} 
Bajaja E., van der Burg G., Faber S.~M., Gallagher J.~S., Knapp G.~R., 
Shane W.~W., 1984, A\&A, 141, 309 

\bibitem[\protect\citeauthoryear{Bajaja et al.}{1988}]{1988A&A...202...35B} 
Bajaja E., Hummel E., Wielebinski R., Dettmar R.-J., 1988, A\&A, 202, 35 
\bibitem[\protect\citeauthoryear{Barnes}{1990}]{1990Natur.344..379B} Barnes 
J.~E., 1990, Natur, 344, 379 
\bibitem[\protect\citeauthoryear{Barnes}{2002}]{2002MNRAS.333..481B} Barnes 
J.~E., 2002, MNRAS, 333, 481 
\bibitem[Bower et al.(2006)]{2006MNRAS.370..645B} Bower, R.~G., Benson, 
A.~J., Malbon, R., Helly, J.~C., Frenk, C.~S., Baugh, C.~M., Cole, S., 
Lacey, C.~G.\ 2006, MNRAS, 370, 645
\bibitem[\protect\citeauthoryear{Buyle et al.}{2006}]{2006ApJ...649..163B} 
Buyle P., Michielsen D., De Rijcke S., Pisano D.~J., Dejonghe H., Freeman 
K., 2006, ApJ, 649, 163 
\bibitem[Caldwell(1984)]{1984ApJ...278...96C} Caldwell, N.\ 1984, ApJ, 278, 
96 
\bibitem[Cole et al.(1994)]{1994MNRAS.271..781C} Cole, S., 
Aragon-Salamanca, A., Frenk, C.~S., Navarro, J.~F., Zepf, S.~E.\ 1994, MNRAS, 271, 781 
\bibitem[Croton et al.(2006)]{2006MNRAS.365...11C} Croton, D.~J., Springel, 
V., White, S.~D.~M., De Lucia, G., Frenk, C.~S., Gao, L., Jenkins, A., 
Kauffmann, G., Navarro, J.~F., Yoshida, N.\ 2006, MNRAS, 365, 11 
\bibitem[\protect\citeauthoryear{De Lucia et 
al.}{2006}]{2006MNRAS.366..499D} De Lucia G., Springel V., White S.~D.~M., 
Croton D., Kauffmann G., 2006, MNRAS, 366, 499 
\bibitem[De Rijcke et al.(2003)]{2003MNRAS.339..225D} De Rijcke, S., 
Zeilinger, W.~W., Dejonghe, H., Hau, G.~K.~T.\ 2003, MNRAS, 339, 225 (DZDH03) 
\bibitem[de Vaucouleurs et al.(1991)]{1991trcb.book.....D} de Vaucouleurs, 
G., de Vaucouleurs, A., Corwin, H.~G., Jr., Buta, R.~J., Paturel, G., 
Fouque, P.\ 1991.\ Third Reference Catalogue of Bright Galaxies.\ Volume 
1-3.~ Springer-Verlag Berlin Heidelberg New York  
\bibitem[\protect\citeauthoryear{Dressler}{1980}]{1980ApJ...236..351D} 
Dressler A., 1980, ApJ, 236, 351 
\bibitem[\protect\citeauthoryear{Driver et al.}{2007}]{2007MNRAS.379.1022D} 
Driver S.~P., Popescu C.~C., Tuffs R.~J., Liske J., Graham A.~W., Allen 
P.~D., de Propris R., 2007, MNRAS, 379, 1022 
\bibitem{}Efstathiou, G., Lake, G., Negroponte, J. 1982 MNRAS, 199, 1069
\bibitem[\protect\citeauthoryear{Emsellem}{1995}]{1995A&A...303..673E} 
Emsellem E., 1995, A\&A, 303, 673 
\bibitem[\protect\citeauthoryear{Emsellem et 
al.}{1996}]{1996A&A...312..777E} Emsellem E., Bacon R., Monnet G., Poulain 
P., 1996, A\&A, 312, 777 
\bibitem[\protect\citeauthoryear{Emsellem \& 
Ferruit}{2000}]{2000A&A...357..111E} Emsellem E., Ferruit P., 2000, A\&A, 
357, 111 
\bibitem[Emsellem et al.(2004)]{2004MNRAS.352..721E} Emsellem, E., and 11 
colleagues 2004, MNRAS, 352, 721 
\bibitem[\protect\citeauthoryear{Emsellem et 
al.}{2007}]{2007MNRAS.379..401E} Emsellem E., et al., 2007, MNRAS, 379, 401 
\bibitem[\protect\citeauthoryear{Flynn et al.}{2006}]{2006MNRAS.372.1149F} 
Flynn C., Holmberg J., Portinari L., Fuchs B., Jahrei{\ss} H., 2006, MNRAS, 
372, 1149 
\bibitem[\protect\citeauthoryear{Garcia}{1993}]{1993A&AS..100...47G} Garcia 
A.~M., 1993, A\&AS, 100, 47 
\bibitem[\protect\citeauthoryear{Georgakakis et 
al.}{2001}]{2001MNRAS.326.1431G} Georgakakis A., Hopkins A.~M., Caulton A., 
Wiklind T., Terlevich A.~I., Forbes D.~A., 2001, MNRAS, 326, 1431 
\bibitem[Gordon et al.(2004)]{2004ApJS..154..215G} Gordon, K.~D., and 28 
colleagues 2004,  ApJS, 154, 215 
\bibitem[\protect\citeauthoryear{Goudfrooij et 
al.}{1994}]{1994A&AS..105..341G} Goudfrooij P., Hansen L., Jorgensen H.~E., 
Norgaard-Nielsen H.~U., 1994, A\&AS, 105, 341 
\bibitem[Gunn and Gott(1972)]{1972ApJ...176....1G} Gunn, J.~E., Gott, 
J.~R.~I.\ 1972.\ On the Infall of Matter Into Clusters of Galaxies and Some 
Effects on Their Evolution.\ Astrophysical Journal 176, 1. 
\bibitem[\protect\citeauthoryear{Hameed \& 
Devereux}{2005}]{2005AJ....129.2597H} Hameed S., Devereux N., 2005, AJ, 
129, 2597 
\bibitem[\protect\citeauthoryear{Hau, Carter, \& 
Balcells}{1999}]{1999MNRAS.306..437H} Hau G.~K.~T., Carter D., Balcells M., 
1999, MNRAS, 306, 437 
\bibitem[Israel(1998)]{1998A&ARv...8..237I} Israel, F.~P.\ 1998.\ Centaurus 
A - NGC 5128.\ Astronomy and Astrophysics Review 8, 237-278. 
\bibitem[\protect\citeauthoryear{J{\'o}zsa et 
al.}{2004}]{2004IAUS..220..177J} J{\'o}zsa G.~I.~G., Oosterloo T.~A., 
Morganti R., Vergani D., 2004, IAUS, 220, 177 
\bibitem[Kauffmann et al.(1993)]{1993MNRAS.264..201K} Kauffmann, G., White, 
S.~D.~M., Guiderdoni, B.\ 1993, MNRAS, 264, 201 
\bibitem[Knapen et al.(1991)]{1991A&A...241...42K} Knapen, J.~H., Hes, R., 
Beckman, J.~E., Peletier, R.~F.\ 1991, A\&A, 241, 42 
\bibitem[\protect\citeauthoryear{Knapp, Turner, \& 
Cunniffe}{1985}]{1985AJ.....90..454K} Knapp G.~R., Turner E.~L., Cunniffe 
P.~E., 1985, AJ, 90, 454 
\bibitem[Kennicutt(1983)]{1983ApJ...272...54K} Kennicutt, R.~C., Jr.\ 
1983, ApJ, 272, 54 
\bibitem[\protect\citeauthoryear{Kennicutt}{1988}]{1988ApJ...334..144K} 
Kennicutt R.~C., Jr., 1988, ApJ, 334, 144 
\bibitem[Kennicutt(1989)]{1989ApJ...344..685K} Kennicutt, R.~C., Jr.\ 
1989, ApJ, 344, 685 
\bibitem[Kennicutt et al.(1989)]{1989ApJ...337..761K} Kennicutt, R.~C., 
Jr., Edgar, B.~K., Hodge, P.~W.\ 1989, ApJ, 337, 761 
\bibitem[\protect\citeauthoryear{Kent}{1988}]{1988AJ.....96..514K} Kent 
S.~M., 1988, AJ, 96, 514 
\bibitem[Kere{\v s} et al.(2005)]{2005MNRAS.363....2K} Kere{\v s}, D., 
Katz, N., Weinberg, D.~H., \& Dav{\'e}, R.\ 2005, MNRAS, 363, 2 
\bibitem[Koo et al.(2005)]{2005ApJS..157..175K} Koo, D.~C., and 14 
colleagues 2005.\ The DEEP Groth Strip Survey. VIII. The Evolution of 
Luminous Field Bulges at Redshift z \~{} 1.\ Astrophysical Journal 
Supplement Series 157, 175-217. 
\bibitem[Kraft et al.(2003)]{2003NewAR..47..625K} Kraft, R.~P., Hardcastle, 
M.~J., Forman, W.~R., Jones, C., Murray, S.~S., Worrall, D.~M.\ 2003.\ High 
resolution X-ray observation and monitoring of the X-ray jet and radio 
lobes of centaurus A.\ New Astronomy Review 47, 625-628. 
\bibitem[Larson et al.(1980)]{1980ApJ...237..692L} Larson, R.~B., Tinsley, 
B.~M., Caldwell, C.~N.\ 1980, ApJ, 237, 692 
\bibitem[Macchetto et al.(1996)]{1996A&AS..120..463M} Macchetto, F., 
Pastoriza, M., Caon, N., Sparks, W.~B., Giavalisco, M., Bender, R., 
Capaccioli, M.\ 1996, A\&AS 120, 463 
\bibitem[\protect\citeauthoryear{Mihos \& 
Hernquist}{1996}]{1996ApJ...464..641M} Mihos J.~C., Hernquist L., 1996, 
ApJ, 464, 641 
\bibitem[Mirabel et al.(1999)]{1999A&A...341..667M} Mirabel, I.~F., 
Laurent, O., Sanders, D.~B., Sauvage, M., Tagger, M., Charmandaris, V., 
Vigroux, L., Gallais, P., Cesarsky, C., Block, D.~L.\ 1999, A\&A, 341, 667 
\bibitem[Morganti et al.(2006)]{2006MNRAS.371..157M} Morganti, R., de 
Zeeuw, P.~T., Oosterloo, T.~A., McDermid, R.~M., Krajnovi{\'c}, D., 
Cappellari, M., Kenn, F., Weijmans, A., Sarzi, M.\ 2006, MNRAS, 371, 157 
\bibitem[Navarro(2004)]{2004ASSL..319..655N} Navarro, J.~F.\ 2004.\ The 
Hierarchical Formation of the Galactic Disk.\ Penetrating Bars Through 
Masks of Cosmic Dust 319, 655. 
\bibitem[Oosterloo et al.(2002)]{2002AJ....123..729O} Oosterloo, T.~A., 
Morganti, R., Sadler, E.~M., Vergani, D., Caldwell, N.\ 2002,  AJ, 123, 729 (O02) 
\bibitem[\protect\citeauthoryear{Oosterloo et 
al.}{2007}]{2007A&A...465..787O} Oosterloo T.~A., Morganti R., Sadler 
E.~M., van der Hulst T., Serra P., 2007, A\&A, 465, 787 
\bibitem[\protect\citeauthoryear{Peng et al.}{2002}]{2002AJ....124..266P} 
Peng C.~Y., Ho L.~C., Impey C.~D., Rix H.-W., 2002, AJ, 124, 266 
\bibitem[\protect\citeauthoryear{Peng, Ford, \& 
Freeman}{2004}]{2004ApJ...602..705P} Peng E.~W., Ford H.~C., Freeman K.~C., 
2004, ApJ, 602, 705 
\bibitem[Phillips et al.(1986)]{1986AJ.....91.1062P} Phillips, M.~M., 
Jenkins, C.~R., Dopita, M.~A., Sadler, E.~M., Binette, L.\ 1986, AJ, 91, 1062 
\bibitem[Reda et al.(2004)]{2004MNRAS.354..851R} Reda, F.~M., Forbes, 
D.~A., Beasley, M.~A., O'Sullivan, E.~J., \& Goudfrooij, P.\ 2004, MNRAS, 
354, 851 
\bibitem[Rejkuba et al.(2002)]{2002ApJ...564..688R} Rejkuba, M., Minniti, 
D., Courbin, F., Silva, D.~R.\ 2002.\ Radio-Optical Alignment and Recent 
Star Formation Associated with Ionized Filaments in the Halo of NGC 5128 
(Centaurus A).\ Astrophysical Journal 564, 688-695. 
\bibitem[\protect\citeauthoryear{Robertson et 
al.}{2006}]{2006ApJ...645..986R} Robertson B., Bullock J.~S., Cox T.~J., Di 
Matteo T., Hernquist L., Springel V., Yoshida N., 2006, ApJ, 645, 986 
\bibitem[\protect\citeauthoryear{Sadler et al.}{2000}]{2000AJ....119.1180S} 
Sadler E.~M., Oosterloo T.~A., Morganti R., Karakas A., 2000, AJ, 119, 1180 
\bibitem[\protect\citeauthoryear{Salim et al.}{2005}]{2005ApJ...619L..39S} 
Salim S., et al., 2005, ApJ, 619, L39 
\bibitem[Schlegel et al.(1998)]{1998ApJ...500..525S} Schlegel, D.~J., 
Finkbeiner, D.~P., Davis, M.\ 1998, ApJ, 500, 525. 
\bibitem[\protect\citeauthoryear{Schiminovich et 
al.}{1994}]{1994ApJ...423L.101S} Schiminovich D., van Gorkom J.~H., van der 
Hulst J.~M., Kasow S., 1994, ApJ, 423, L101 
\bibitem[Scorza et al.(1998)]{1998A&AS..131..265S} Scorza, C., Bender, R., 
Winkelmann, C., Capaccioli, M., Macchetto, D.~F.\ 1998, A\&AS 131, 265
\bibitem[Somerville and Primack(1999)]{1999MNRAS.310.1087S} Somerville, 
R.~S., Primack, J.~R.\ 1999, MNRAS, 310, 1087
\bibitem[Telesco(1978)]{1978ApJ...226L.125T} Telesco, C.~M.\ 1978.\ 
Extended 10 micron emission from the dark lane in NGC 5128 /Centaurus A/.\ 
Astrophysical Journal 226, L125-L128. 
\bibitem[\protect\citeauthoryear{Terlevich \& 
Forbes}{2002}]{2002MNRAS.330..547T} Terlevich A.~I., Forbes D.~A., 2002, 
MNRAS, 330, 547 
\bibitem[Tinsley(1980)]{1980FCPh....5..287T} Tinsley, B.~M.\ 1980.\ 
Evolution of the Stars and Gas in Galaxies.\ Fundamentals of Cosmic Physics 
5, 287-388. 
\bibitem[\protect\citeauthoryear{Toomre}{1964}]{1964ApJ...139.1217T} Toomre 
A., 1964, ApJ, 139, 1217 
\bibitem[Toomre and Toomre(1972)]{1972ApJ...178..623T} Toomre, A., Toomre, 
J.\ 1972, ApJ, 178, 623 
\bibitem[\protect\citeauthoryear{Trager et al.}{2000}]{2000AJ....120..165T} 
Trager S.~C., Faber S.~M., Worthey G., Gonz{\'a}lez J.~J., 2000, AJ, 120, 
165 
\bibitem[\protect\citeauthoryear{van der Burg \& 
Shane}{1986}]{1986A&A...168...49V} van der Burg G., Shane W.~W., 1986, 
A\&A, 168, 49 
\bibitem[\protect\citeauthoryear{van Gorkom et 
al.}{1990}]{1990AJ.....99.1781V} van Gorkom J.~H., van der Hulst J.~M., 
Haschick A.~D., Tubbs A.~D., 1990, AJ, 99, 1781 
\bibitem[\protect\citeauthoryear{Wilson \& 
Scoville}{1992}]{1992ApJ...385..512W} Wilson C.~D., Scoville N., 1992, ApJ, 
385, 512 
\bibitem[\protect\citeauthoryear{Wilson \& 
Matthews}{1995}]{1995ApJ...455..125W} Wilson C.~D., Matthews B.~C., 1995, 
ApJ, 455, 125 
\bibitem[White and Frenk(1991)]{1991ApJ...379...52W} White, S.~D.~M., 
Frenk, C.~S.\ 1991, ApJ, 379, 52 
\item[]Worthey, G., 1994, ApJS, 95, 107.
\bibitem[\protect\citeauthoryear{Yi et al.}{2005}]{2005ApJ...619L.111Y} Yi 
S.~K., et al., 2005, ApJ, 619, L111 


\end{thebibliography}
\end{document}